# An Approach of Digital Image Copyright Protection by Using Watermarking Technology

Md. Selim Reza[1], Mohammed Shafiul Alam Khan[1], Md. Golam Robiul Alam[2], and Serajul Islam[1]

[1] Institute of Information Technology, University of Dhaka, Dhaka-1000, Bangladesh
selimr_ict@yahoo.com, shafiul@univdhaka.edu, nirobtalukder@gmail.com

[2] International Islamic University Chittagong, Chittagong-4203, Bangladesh
gra9710@yahoo.com

**Abstract**

Digital watermarking system is a paramount for safeguarding valuable resources and information. Digital watermarks are generally imperceptible to the human eye and ear. Digital watermark can be used in video, audio and digital images for a wide variety of applications such as copy prevention right management, authentication and filtering of internet content. The proposed system is able to protect copyright or owner identification of digital media, such as audio, image, video, or text. The system permutated the watermark and embed the permutated watermark into the wavelet coefficients of the original image by using a key. The key is randomly generated and used to select the locations in the wavelet domain in which to embed the permutated watermark. Finally, the system combines the concept of cryptography and digital watermarking techniques to implement a more secure digital watermarking system.

***Keywords:*** *Copyright Protection, Digital Watermarking, Discrete Wavelet Transform, Watermark Extraction.*

## 1. Introduction

The recent digital culture provides many new opportunities for the rapid and inexpensive distribution of digital content. Various kinds of information can be encoded into digital form, duplicated without loss of fidelity, and transmitted to incredible numbers of recipients worldwide at negligible cost. While offering interesting opportunities, this evolution also poses serious challenges in enforcing intellectual property rights, as effortless duplication and distribution inevitably leads to illegal circulation and usage of copyrighted material. Copyright protection of digital image is defined as the process of proving the intellectual property rights to a court of law against the unauthorized reproduction, processing, transformation or broadcasting of a digital image. Depending on the law in various countries, the process enable on a prior registration of the copyright with a trusted third party. After successful registration, the copyright ownership legally bound by a copyright notice, which is required to notify and prove copyright ownership [7].

Digital watermarking is a method to make data set, such as image, sound or video. A stego data set consist of the original data, the cover data set and a digital watermark that does not affect the data set's usability but that can be detected using dedicated software or system. Watermarking can be used for marking authorship or ownership of a data set [7].

This paper focuses on the problems of current copy protection technologies. The focus of the paper is at embedding a binary or gray-scale digital image into the cover media. The embedding depends on the wavelet coefficient of the original image under consideration. The manipulation of the image pixels based on the limitation of the human eye to detect slight changes in gray levels.

## 2. Literature Review

Digital watermarking is used to embed some information into the digital media which needs to be protected from illegal copying and for the purpose of authentication. It is the digital counter-part of paper water-marking. A trademark, a seal or a sequence number is usually embedded. Digital watermarking must be robust, that is it must have the ability to survive common image processing operations, such as lossy compression, filtering, noise adding, geometrical transformation, etc. Watermarks are embedded directly into a file data, usually by making minor variations to pixel brightness. The variations in the data bits are slight and cannot be detected by the human eye. The patterns are repeated many times, allowing robustness. In addition, it must also be invisible and unambiguous. Verification of existence of watermarking in the absence of original media is also an essential requirement.

2.1 Types of Digital Watermarking

Based on the type of document to be watermarked [4][8]

**Image Watermarking**: A watermark is embedding into the image by using different techniques, such as spatial domain watermarking, frequency domain watermarking etc.

**Video Watermarking**: It is the process of watermarking the sequence of video frames. One can watermark the raw frame data, or the compressed data.

**Audio Watermarking**: Audio, like any other data, constitutes a carrier for physo-acoustically hidden data. Thus the process of embedding information into audio can be termed as audio watermarking.

**Hardware/Software Watermarking**: Software watermarking is used to hide customer identification or copyright information within software applications. A watermark is used to identify owners of the software or to stealthily identify the origin of a pirated copy. In the hardware context, Boolean equivalences can be exploited to yield instances that use different types of gates and that can be addressed by the hidden information bits.

**Text Watermarking**: This is done at two levels. At the printout level, information can be encoded in the way the text lines or words are separated. At the semantic level (necessary when raw text files are provided), equivalences between words or expressions can be used.

Based on the human perception [4][8]

**Visible Watermark**: A visible watermark is a visible semi-transparent text or image overlaid on the original image. It allows the original image to be viewed, but it still provides copyright protection by marking the image as its owner's property. We see visible watermarks every day watching television, that is, TV station logos.

**Invisible Watermark**: It can be classified in two categories, Invisible-robust watermark and invisible-fragile watermark. Invisible-robust watermark is embedded in such a way that an alteration made to the pixel value is perceptually not noticed and it can be recovered only with appropriate decoding mechanism. The invisible-fragile watermark is embedded in such a way that any manipulation or modification of the image would alter or destroy the watermark.

Dual watermark is a combination of a visible and an invisible watermark. In this type of watermark, an invisible watermark is used as a backup for the visible watermark.

The requirements that a watermarking system needs to comply with depends upon the specific type of application. A few most common applications involve such as copyright protection, copy protection, broadcast monitoring, medical applications, fingerprinting, data authentication etc [3][4].

For the copyright protection, a digital watermarking technique must satisfy the requirements like perceptual transparency, robustness, universality, capacity, and payload unambiguousness tamper-resistance [3][4][6].

2.2 Domain of Digital Watermarking

The domain of digital watermarking can be of two categories:

**Spatial Domain Watermarking:** Spatial domain watermarking slightly modifies the pixels of one or two randomly selected subsets of an image. Modification might include flipping the low-order bit of each pixel. However, this technique is not reliable when subjected to normal media operation such as filtering or lossy compression.

**Frequency Domain Watermarking:** This technique is also known as transforming domain. Values of certain frequencies are altered from their original. Typically, these frequency alterations are done in the lower frequency levels, since alternations at the higher frequencies are lost during compression. The watermark is applied to the whole image so as not to be removed during a cropping operation.

2.3 Discrete Wavelet Transform

It is one of the frequency transformation techniques used in this paper. Wavelets are functions that satisfy certain mathematical requirements and are used in representing data or other functions. The idea is not new. Approximation using superposition of functions has existed since early 1800's, when Joseph Fourier discovered that he could superpose sine and cosine to represent other functions. However, in wavelet analysis, the scale that we use to look at data plays a special role.

Wavelet algorithms process data at different scales and resolutions. If we look at a signal with a large "window", we would notice gross features. Similarly, if we look at a signal with a small "window", we would notice small features. The result in wavelet analysis is to see both the forest and the trees. The wavelet transform is a linear transform that generalizes the properties of the Haar transform. A wavelet, in the sense of the Discrete Wavelet Transform (DWT), is an orthogonal function which can be applied to a finite group of data. Wavelets are functions that satisfy certain requirements localized in time and frequency; integrate to zero, quick and easy calculation of the direct and inverse wavelet transformation [1][2][4][5][9][10].

In wavelet analysis, an original image can be decomposed into an approximate image LL and three detail images LH, HL and HH. Using wavelet analysis on the approximate image again, four lower-resolution sub-band images will be got, and among them, the approximate image holds most of the information of the original image.

The DWT of an image has two parts, an approximation part (this is an image with smaller dimensions) and a detail part (this is a set of images with smaller dimensions containing the details of the original image).

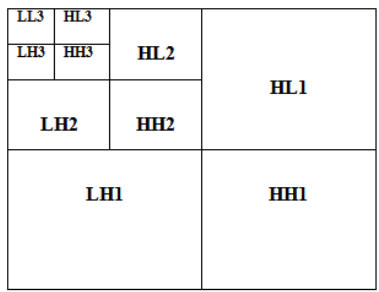

Fig. 1 DWT pyramid decomposition of an image

Hence, the DWT gives access to the detail of the original image. This is very important because changing only the less important detail of an image is easy to insert a watermark in this image, keeping the insertion procedure invisible. Three-level wavelet decomposition with gray area indicating highest frequency component, black area indicating lowest-frequency component, and white areas indicating the hiding places. Where, H and L mean the high pass and low pass filter, respectively. While HH means that the high pass filter is applied to signals of both directions [1][2][4][5].

## 3. Existing Digital Watermarking Technologies

In this section techniques based on wavelet domain and permutation approaches are discussed.

### 3.1 Copyright Protection using Watermarking Algorithm [1]

This method was proposed by Abou Ella Hassanien. In this paper, a robust image watermarking algorithm for copyright protection based on two-dimensional discrete wavelet transform using Human Visual System (HVS) is introduced. To make watermark robust we embed the watermark in the higher level sub-band (but not in the scaling coefficients), even though it may affect the perceptual invisibility of the image. By careful embedding the watermark, it will not cause much change in the image fidelity. The wavelet transform breaks an image down into four sub-sampled images. The result consist of one image that has high pass in the horizontal and vertical directions, one that has low passed in the vertical and high passed in the horizontal, and one that has low pass filtered in both directions.

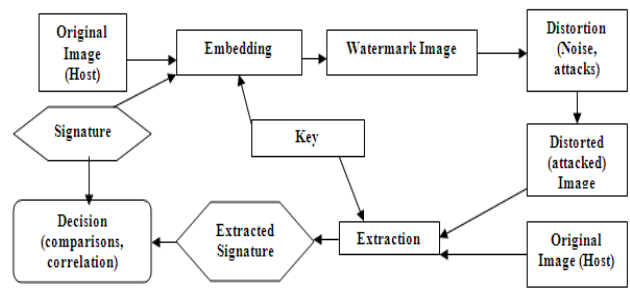

Fig. 2 Block diagram of digital watermarking in wavelet domain

A new metric that measures the objective quality of the image based on the detected watermark bit is introduced. The original unmarked image is not required for watermark extraction. The performance of the proposed watermarking algorithm is robust to variety of signal distortions, such a JPEG, image cropping, geometric transformations and noises.

### 3.2 An Adaptive Digital Image Watermarking Technique for Copyright Protection [6]

This technique was proposed by Chang-Hsing Lee and Yeuan-Kuen Lee. In this paper, the watermark is used as a binary image. That's why human eye can easily identify the extracted watermark. In fact, embedding a watermark in the least significant bits of a pixel is less sensitive to human eyes. However, the watermark will be destroyed if some common image operations such as low-pass filtering are applied to the watermarked image. Therefore, to make the embedded watermark more resistant to any attack, the watermark must be embedded in the more significant bits. This will introduce more distortion to the host image and conflicts with the invisible requirement. To meet both invisible and robust requirements, modify the intensities of some selected pixels as large as possible and this modification is not noticeable to human eyes. In addition, to prevent unauthorized access, the watermark is first permutated into scrambled data. The block diagram of the watermark system is depicted in Fig. 3

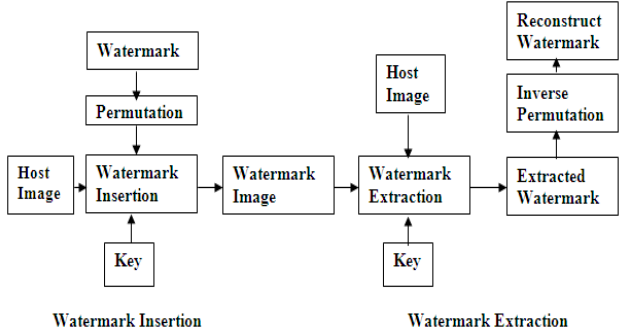

Fig. 3 Block diagram of permutation based watermarking

# 4. Proposed Digital Watermark Method

In our proposed method, discrete wavelet transform and permutation technique are used to distort the original image and watermark, respectively to embed watermark into the original image. We use a binary or gray-scale logo as a watermark. The permutated watermark embeds in the wavelet coefficients of the DWT of an image.

Assuming that the watermark of length $Nw$ is binary and consists of elements from the set {0, 1}. To embed the mark in the discrete wavelet domain, relative changes are performed on the coefficients so that the host image is not required for extraction. We first permutated the watermark and then embed it into the detail components of the original image with the use of a key.

This key $k$ consists of three components:

a. The particular DWT used to embed the watermark
b. The embedding parameter $Q$
c. The coefficient selection key $Ks$

The coefficient selection key $Ks$ is randomly generated by the Congruntial algorithm that is used to select the exact locations in the wavelet domain in which to embed the permutated watermark. For each coefficient within the wavelet domain the key $Ks$, has a value of one or zero to denote that the coefficient is to mark or not, respectively.

## 4.1 Key Generation Algorithm

a. Start with a seed value $Z0$ and length $L$
b. Calculate the $Zi = (a*Zi-1+c)\ mod\ m$ where $a$ = multipler, $c$ = increment and $m$ = modulus
c. $Ks(i) = Zi/m$;
d. Increment $c = c+1$ and $i = i+1$ where $i$ = index of $Ks$
e. Continue step 2 until $c = L$

## 4.2 Watermark Embedding

In this phase, at first we apply the permutation on the watermark to scramble the watermark for which the watermark is invisible to the human visual system. We embed the permutated watermark into the detail wavelet coefficients of the original image with the use of key. However, in the wavelet base watermarking the watermark is spread all over the image. The block diagram of our proposed method for watermark embedding is shown in Fig. 4.

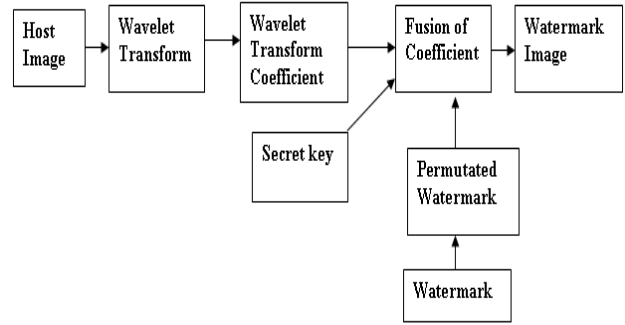

Fig. 4 Block diagram of the proposed watermark embedding

**Step 1:** The original image is transformed into the discrete wavelet domain. Specifically, we perform the $L^{th}$ level DWT of the original image to produce a sequence of $3L$ detail images at each of the L resolution levels, and a gross approximation of the image at the coarsest resolution level. We denote the $O^{th}$ frequency orientation at the $l^{th}$ resolution level of the image $f$ by $f_{o,l}(m,n)$ where $O \in \{1,2,3\}$ represents the frequency orientation corresponding to the horizontal, diagonal and vertical image details, $l \in \{1,2, . . , L\}$ is the resolution level and $(m, n)$ is the particular spatial location index at the resolution 1. The gross approximation is represented by $f_{4,L}(m, n)$ where the subscript "4" is used instead of $O$ to denote the gross image approximation at resolution $L$.

**Step 2:** In this step we permutated the watermark and the permutated watermark bits are embedded in the original signal coefficients specified by the key *Ks*. If Ks specifies that the location *(m, n)* at resolution l can be marked, then we perform the following steps:

1. The detail coefficients *f1,l(m, n), f2,l(m, n)* and *f3,l(m, n)* are sorted in ascending order. We denote these ordered coefficients by *fo1,l(m, n), fo2,l(m, n)* and *fo3,l(m, n)* where,

$fo1,l(m, n) \leq fo2,l(m, n) \leq fo3,l(m, n)$

Such that *o1, o2, o3* $\in$ {1, 2, 3) and *o1 $\neq$ o2, o2 $\neq$ o3:* and *o1 $\neq$ o3*.

2. One permutated watermark bit is embedded by modifying the median value of the detail coefficients at resolution *l* (i.e., *fo2,l(m,n)*) at spatial location *(m,n)*. To embed the watermark, we quantize *fo2,l(m,n)*. The range of values between *fo1,l(m,n)* and *fo3,l (m,n)* are divided into bins of width

$$\delta = \frac{fo3,l(m, n) - fo1,l(m,n)}{2Q-1} \quad (1)$$

where *Q* is a key-specified quantization variable. To embed a permutated watermark bit of value zero or one *fo2,l (m, n)* is quantized to the nearest value. The new watermarked coefficients are denoted *fwo,l (m, n)*.

It should be explained that an attacker cannot easily determine the exact key *Ks*, and remove the watermark given only the watermarked image if the specific wavelet transform used in the decomposition of stage *I* is kept secret and *Q* is unknown.

Therefore, it is not possible to use the relative value of the coefficients to determine the watermark locations and hence destroy the mark by randomly changing the coefficient dues by small amounts. The value of *Q* determines the trade-off between robustness and visibility of the watermark. The smaller its value, the more robust is the mark.

**Step 3:** The corresponding $L^{th}$ level inverse DWT of the coefficients *fwo,l(m,n)* is computed to form the watermarked image *fw*. The complexity of the first and third stages depends on the particular DWT employed (which is specified in *K*) and its associated implementation.

### 4.3 Watermark Extraction

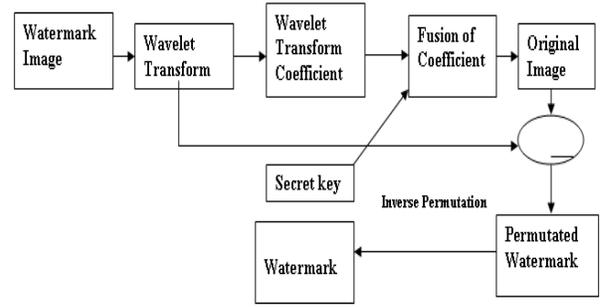

Fig. 5 Block diagram of Watermark Extraction

Let *f'w*, be the signal from which we would like to extract the watermark. We first apply an $L^{th}$ level DWT on *f'w* to produce the coefficients *fwo,l'(m,n)*. We use the key *Ks*, to find the locations in which the watermark was embedded for each resolution level *l*. We extract the watermark bits from these coefficients as follows:

1. The detail coefficients *f1,l(m, n), f2,l(m, n)* and *f3,l(m, n)* are sorted in ascending order. We represent the ordered coefficients by

*fw'o1,l(m, n)* $\leq$ *fw'o2,l (m, n)* $\leq$ *fw'o3,l (m, n)*

Such that *o1, o2, o3* $\in$ *{1, 2, 3)* and o1 $\neq$ o2, o2 $\neq$ o3 and o1 $\neq$ o3.

2. The watermark bit value is estimated from the relative position of *fw'o2,l (m, n)*. Using the same value of *Q* as for embedding (which is given by *K*), the watermark value is determined by finding the closest quantized value, to *fw'o2,l (m,n)* and converting it to its associated binary number.

3. If the watermark sequence has been embedded several times, then the most common extracted bit value is taken for the watermark estimate.

Our approach increased performance from characterizing the distortion on the watermarked signal for fusion watermarking. We propose a method to improve the performance of a broad class of watermarking schemes through attack characterization.

### 5. Implementation

There are different languages to implementing any project like C, C++, Java, Matlab etc. We have chosen the Matlab v7.1.0.246 (R14) Service Pack 3 of Mathworks Incorporation due to its vast collection of computational algorithms and mathematical functions. Many built in functions for image processing in MATLAB's are easier than other languages which are needed in the implementation of our algorithm.

Inputs:

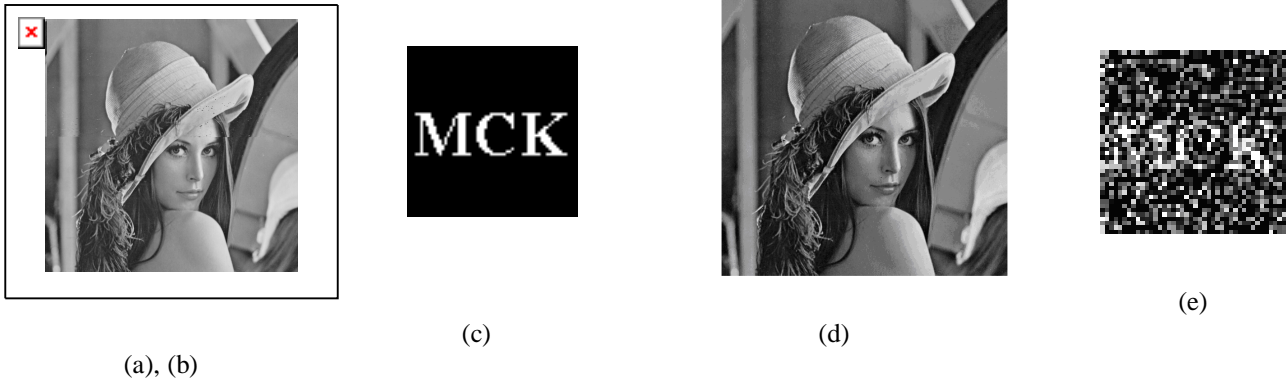

(a), (b)            (c)            (d)            (e)

Fig. 6 (a) Original Image of size 512*512, (b) 4-level Wavelet Decomposition of Original Image, (c) Watermark (d) Watermarked Image (e) Extracted Watermark

## 6. Comparative Study

In this section, comparisons of existing methods to proposed method are presented. In Abou Ella Hassanien's method, the key is randomly generated from the image and takes more time for generating the key. Chang -Hsing Lee and Yeuan-Kuen Lee's method, the watermark used as a key rather than a randomly generated number. Proposed method uses the congruntial algorithm for generating the key based on a seed value. The key is generated by using following forms, *Ks(j): = randSeed (0,1)* which is based on a uniform random number generator between 0 and 1.

In the embedding stage Abou Ella Hassanien's method, decompose the original image into the 3-level wavelet decomposition. The watermark values are repeatedly embedded in different coefficients selected by the key. Chang-Hsing Lee and Yeuan-Kuen Lee's method, permutated watermark embed into the spatial location of the host image.

In proposed method, we decompose the original image into 4-level wavelet transform using Daubechies-4 wavelet and also permutated the watermark. Permutated watermark values are repeatedly embedded in different coefficients selected by the key so that it will be able to spread the watermark to all over the original image and more robust to attacks on embedded watermark.

In the extraction phase, Abou Ella Hassanien's method, need the watermark and key for extraction. Chang-Hsing Lee and Yeuan-Kuen Lee's method need original image for watermark extraction. Our proposed method can extract the watermark without reference to original image, only need the key.

## 7. Conclusion

In this paper, a robust image watermarking approaches for copyright protection based on discrete wavelet transform and permutation is proposed. The steps of the proposed algorithm, including watermark embedding and watermark detection is described. By using wavelet, we spread the watermark all over the image and by permutation watermark is invisible to the human eye. The proposed method is robust to various attacks, such as image compression, image filtering, geometric transformation and noises.

Digital watermark of still images based on the wavelet transform and permutation is a good method for authentication of image materials. The using of wavelet transform satisfies requirements for watermarked media. We demonstrate the investigation of image distortion on relation between the original image and watermarked image.

In future we can apply this technique in audio, video or other media objects and we can study the state-of-the-art speech and speaker recognition techniques, and to embed the speaker and speech content in the audio signal. There is also an idea to study how to embed the watermark into the video frames referred as a video watermarking.

**Md. Selim Reza** was born in Tangail, Bangladesh. He received his B. Sc. Engineering in ICT from Mawlana Bashani Science and Technology University (MBSTU), Tangail, Bangladesh in 2007 and M. Sc. in Information Technology from the University of Dhaka, Bangladesh in 2011. He is working as an Assistant Programmer in Basic ICT Skill Transfer up to Upozilla Level Project, BCC, Under Ministry of ICT, Bangladesh. His current research interests include Networking (with Green Wireless Networking), Communication, Digital Content Security, Image processing, Embedded System, Re-configurable Computing and Software Development.

**Mohammed Shafiul Alam Khan** was born in Mymensingh, Bangladesh. He received B. Sc. in Computer Science and Engineering (CSE) from University of Dhaka in 2005 and MS in CSE from University of Dhaka in 2007. He is now Assistant Professor at Institute of Information Technology (IIT) in University of Dhaka. His research interests include Data / Network Security, Wireless Networks, Data Mining and Information Security. He is a co-author of about fifteen research articles in different journals and conferences.

**Md Golam Rabiul Alam** was born in Rangpur, Bangladesh. He received the Bachelor Degree in Computer Science & Engineering from Khulna University in 2002 and the Master Degree in Information Technology from University of Dhaka in 2011. He is now an Assistant Professor of Department of Computer Science and Engineering in International Islamic University Chittagong (IIUC). His current research interests include the Heterogeneous Wireless Networks, Data Mining and Information Security. He is a co-author of about thirty research papers and a textbook.